\documentclass[aps,prl,twocolumn,showpacs]{revtex4}
\usepackage{amsmath}
\usepackage{epsfig}
\usepackage{amssymb}
\usepackage{dcolumn}
\usepackage{graphicx}
\usepackage{dcolumn}

\usepackage{color}

\newcommand{\beq}{\begin{equation}}
\newcommand{\eeq}{\end{equation}}

\begin{document}
\date{\today}

\title{Optical computing with soliton trains in Bose-Einstein condensates}

\author{Florian Pinsker}
\email{fp278@cam.ac.uk}

\affiliation{Department of Applied Mathematics and 
Theoretical Physics, University of Cambridge, United Kingdom.}

\begin{abstract}
Optical computing devices can be implemented based on controlled generation of soliton trains in single and multicomponent Bose-Einstein condensates (BEC). Our concepts utilize the phenomenon that the frequency of soliton trains in BEC can be governed by changing interactions within the atom cloud \cite{dark}. We use this property to store numbers in terms of those frequencies for a short time until observation.   The properties of soliton trains can be changed in an intended way by other components of BEC occupying comparable states or via phase engineering. We elucidate in which sense such an additional degree of freedom can be regarded as a tool for controlled manipulation of data. Finally the outcome of any manipulation made is read out by observing the signature within the density profile.

\end{abstract}


\maketitle

\section{Introduction} Since the first observation of Bose-Einstein condensation in weakly-interacting and dilute Bose gases at low temperatures \cite{first,second,third}, BEC have become an exciting playground for scientists to investigate their characteristic properties \cite{app,leg,dilute,LSY,non}. The concept of macroscopically populated states in phase space \cite{E,OP} led to condensates consisting of photons  \cite{photon,Light}, of quasiparticles such as exciton-polaritons \cite{exit} and more recently of classical waves \cite{Fleisch}. Generating dilute atomic BEC experimentally was first accomplished $70$ years after the proposed existence \cite{ket2, E}. More recently the experimental setups could be miniaturized to small chips \cite{chip,chip2} and our command over BEC has by now become highly sophisticated \cite{leg,app}. The purpose of this letter is to propose an application for BEC as certain {\it classical computing devices} .

A key characteristic for our concept is that of dimension reduction in highly tightened cigar-shaped traps to effectively $1d$ condensates \cite{marko,ket1}.   Those systems obey stable nonlinear excitations, such as solitons \cite{reviews}. Solitons have been observed experimentally both within BEC with attractive and repulsive interactions \cite{dbec,carr}. In motion solitons move with constant speed and under particular circumstances form arrays, so called soliton trains \cite{nat, dark}. Feshbach resonances can be utilized to manipulate scattering lengths \cite{nat, fesch}  locally and in particular enable the experimenter to {\it generate} nonlinear excitations \cite{BadPan} such as {\it soliton trains}  in a highly controllable way \cite{dark}. Another effective approach to induce solitons into BEC are phase imprinting methods \cite{dbec, pow, burg, burg2}.  Furthermore the existence of various stable excitations in multicomponent atomic BEC has  been reported for systems of two components in similar states \cite{vrings}. A recent study has shown the effective creation of soliton trains in one-dimensional single and multicomponent BEC by changing interactions within the condensate on one side of the confining trap \cite{dark} or in exciton-polariton condensates by applying a proper potential step at a gaussian pumping spot \cite{cool}. For atomic condensates the change in interactions $s$ is uniquely related to the frequencies $f$ of emerging soliton trains.  In this letter we propose to utilize such an association for storing numbers for a short time, which are to be calculated by the presence of other components of BEC in comparable states or via phase imprinting techniques. The outcome of any calculation is received by measuring the density profile of the  condensate. Our scheme essentially relies on the measurability of soliton trains, which have been observed in experiments for a variety of settings \cite{dbec, carr,bing,bing2,alex,PRA,non}. Solitons could be observed in a wide range of different physical situations among which optical solitons \cite{Kiv} are widely used in today's technology \cite{classic2}. For several decades light waves have been utilized for a wide range of applications such as in nonlinear fibre optic communication \cite{classic, classic2, classic3, classic4} while research on new technologies is thriving in particular on elementary circuit components such as diodes \cite{diode} or transistors \cite{trans} utilizing light or realized in polariton condensates \cite{trans22,trans33} and conceptually on optical computing schemes \cite{opco}.

\section{  Mathematical model} Our concept is developed within the theoretical framework of Gross-Pitaevskii theory for  $N$ distinguishable components of atomic BEC in highly elongated cigar shaped traps \cite{reviews, marko}. Formally we consider a system of effectively onedimensional coupled nonlinear Schr\"odinger-type equations in nondimensionalized form, 
\begin{equation}\label{multi}
 i \frac{\partial}{\partial t}  \psi_{i}= - \frac{\partial^2}{\partial z^2} \psi_{i}  + g_i |\psi_{i}|^2 \psi_{i} + \sum^N_{j \neq i} g_{ij} |\psi_{j}|^2 \psi_{i}
\end{equation} 
with $i,j \in \mathbb N^+$ and $i \neq j$. Here $\psi_i = \psi_i (z)$ with $z \in \mathbb R$ represents the condensate wave function of component $i$, $g_i$ denotes the corresponding self-interaction strength and $g_{ij}$ cross-interactions between component $i$ and $j$. We neglect an external trapping potential as we consider highly elongated condensates. Very loose confinement in $z$-direction is necessary to allow soliton trains to emerge \cite{dark} as for tighter taps solitons get reflected on its boundaries \cite{konotop}. All Bosons of all components are assumed to be identically and uniformly distributed initially, i.e., $\psi_i = \sqrt{n_0}$, where $n_0$ represents the constant number density distribution of Bosons for each component. We refer to \cite{dark, reviews, marko} for details on onedimensional Gross-Pitaevskii theory of atomic gases, but point out that our proposal is universally applicable for all systems described by $\eqref{multi}$ satisfying the following assumptions.
Two types of computing schemes are to be discussed independently based on two different sets of  {\it necessary properties} of the underlying  physical system:

Method A:
$1.$ We consider multicomponent systems ($N>1$) and allow cross- and self-interactions to be step-like functions in space. 
$2.$ Different components can occupy states such that their density distributions satisfy $|\psi_i|^2 \simeq |\psi_j|^2$.

Method B: $1.$ We consider a single condensate ($N=1$). Self-interactions can be changed to become a step-like function.
$2.$ There exists a mechanism to locally change the phase of the condensate wave function over time.

\section{  Working memory} 
\begin{figure}
\epsfig{file=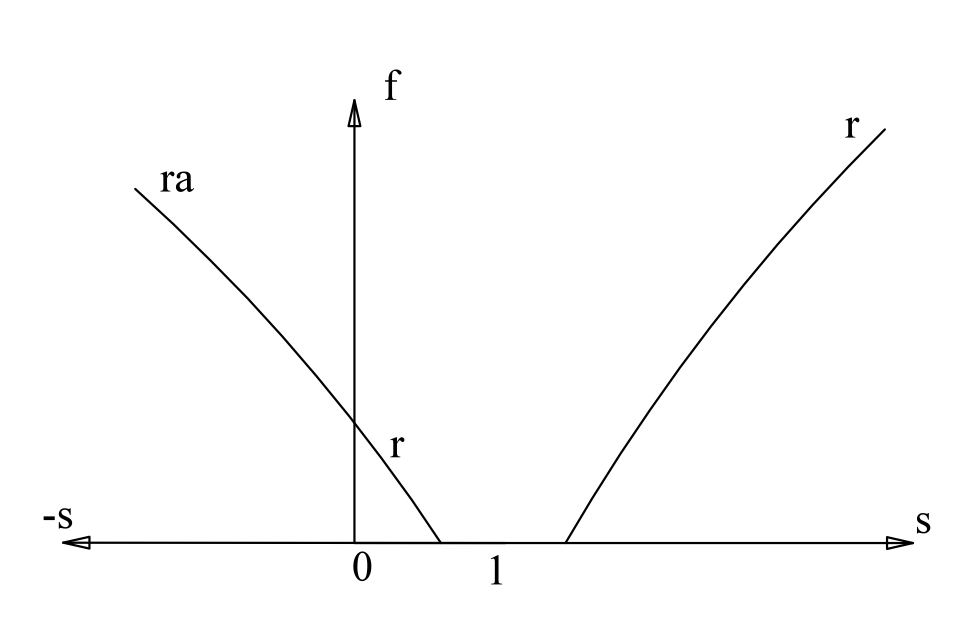 ,width = 1. \columnwidth}
\caption{Frequency of the soliton train as a function of change in interatomic interactions. $r$ denotes the frequency due to a change of interactions within a repulsive condensate and $ra$ due to changing to attractive interactions on one part of the condensate of an initially repulsive condensate \cite{dark} \label{freq}. }
\end{figure}For each single component BEC $\psi_i$ with repulsive and constant self-interactions and without cross-interactions, i.e.,  $g_i(z) =g_i$ and $g_{ij} = 0$ $ \forall i,j$ imposing an action such that interactions become step-function like, i.e., formally
\beq\label{step}
 g_i = c_1  \to g_i(z) =  \begin{cases} g^R=c_1, & z>0  \\ g^L =c_2, & z \leq 0, \end{cases}
\eeq
enables soliton trains to emerge \cite{dark}. After such change is implemented the left hand side ($z \leq 0$) of the condensate has self-interactions $g^L=c_2$ and the right hand side remains to obey interactions $g^R=c_1$, where $c_1$ and $c_2$ denote fixed real-valued constants.  The relationship between change in interactions $s = g^L/g^R$ and frequency of the generated soliton train is illustrated in Fig. $\ref{freq}$ and represented by the function $r(s)$. We note that theoretically soliton trains occur above a threshold of change  given by $2.2$ or below $1/2.2$ in massive infinitely spread repulsive BEC at zero temperature \cite{dark}. For atomic BEC in the Gross-Pitaevskii regime this critical value, however, might as well depend on the fraction between BEC and the thermalized atom cloud present in experiments and so we just write $c^{\text{up}}_{\text{crit}}$ for the critical value for which increased interactions and  $c^{\text{down}}_{\text{crit}}$ for which decreased interactions lead to generation of soliton trains.
The association of change in interatomic interactions $s \in C \subset \mathbb R^{c^{\text{up}}_{\text{crit}}}$  with the frequency of dark solitons to emerge from such action $f \in F \subset \mathbb R^+$ represents a bijective map between both properties, which in addition is monotonic increasing for $s >c^{\text{up}}_{\text{crit}}$ (otherwise decreasing) and continuous as Fig. \ref{freq} illustrates. Hence, this physical system constitutes a data storage in so far as one can imprint some number $a$ into the BEC by first associating it with the amount of change in self-interactions $s(a)$ bijectively. The change in interactions imposed on the physical system then leads to a dark soliton train of certain frequency $f(s(a))$, i.e., each number has its frequency associated uniquely (for each branch in Fig. \ref{freq}).  Moreover the frequency $f(s(a))$ of any soliton train can be read out by measuring the density profile of the condensate. We note that a bijection is sufficient for storing information like a number that can be read out again, as it provides the existence of an inverse function. Therewith by observing the frequency of a soliton train one restores the information put into the condensate via the inverse functions, i.e., $s^{-1}(f^{-1}(f(s(a))))=a$. 

\section{    Computing  method A} For two component systems ($N=2$) we utilize above concept of short term memory by invoking the additional freedom to manipulate the input via cross-interactions. We suppose that both condensates say $1$ and $2$ are in similar states, i.e., $|\psi_1|^2 \simeq |\psi_2|^2$, which simplifies the system of equations $\eqref{multi}$ to a single equation. The effective equation for both states $\psi_i$ with $i \in \{1,2 \}$  is
\begin{equation}\label{two}
 i \frac{\partial}{\partial t}  \psi_{i}= - \frac{\partial^2}{\partial z^2} \psi_{i}  + g^{\rm eff}_{i} |\psi_{i}|^2 \psi_{i},
\end{equation}
where we introduced the effective interaction $g^{\rm eff}_{i} = (g_i + g_{ij}) > 0$, which includes both cross- and self-interactions. The crucial property of such a two component BEC for the task of generating soliton trains of a certain frequency is that cross-interactions can take over the role of self-interactions and vice versa.

\begin{figure}
\epsfig{file=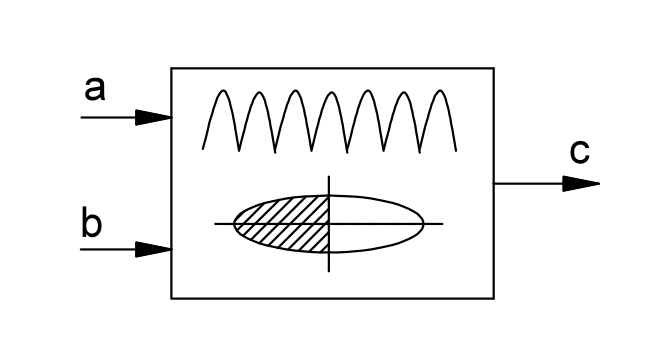 ,width = 0.6\columnwidth}
\caption{Illustration of input number $a$ and $b$ being processed to an output number $c$ in an intended way by means of two distinct mechanisms generating observable soliton trains of  certain frequencies. \label{symb}}
\end{figure}

 If we would imprint soliton trains of a certain frequency in both components by changing self-interactions corresponding bijectively to a number, we could modify the input in each component by the presence of the other component by changing cross-interactions between them as well and at the same instant of time. For example consider the input number $a \in \mathbb R_{+}$ that corresponds to the frequency of the soliton trains generated in components $A$ and $B$ due to a change in self-interactions in both components. We can add theoretically any other number $b \in \mathbb R_{+}$ by turning on cross-interactions between $A$ and $B$ in an appropriate sense, since the frequency of dark soliton trains in component $A$ and $B$ is changed by such a manipulation and now can be associated to a different real number $c$. The concept of the device is illustrated in Fig. \ref{symb}, where $a$ and $b$ represent the input numbers and $c$ the output number. 
Clearly the properties of the new number $c$ are defined by the two-component system used for this computation. By the assumption that both self- and cross-interactions can be changed the {\it  effective } `self-interaction' for each state can be written as
\beq\label{not}
g^{\rm eff}_{i} (z) = g_i (z) + g_{ij} (z)
\eeq
with $g_i(z)>0$,
\beq\label{puss}
g_{ij} = c_3 \to g_{ij}(z) =  \begin{cases} g^R_{ij}=c_3, & z>0  \\ g^L_{ij} =c_4, & z \leq 0, \end{cases}
\eeq
and also $g_{ij}(z)  \geq 0$ while $c_3$ and $c_4$ are constants. More specifically we consider first cross-interactions of the form $g^R_{ij} = 0$ and $g^L_{ij}   \geq 0$. Here we assume initial cross-interactions to be turned off and after a change is induced cross-interactions on the r.h.s. stay turned off  ($g^R_{ij} = 0$), but a change on the l.h.s. $g^L_{ij} \geq 0$ is implemented. 
 \begin{figure}
\epsfig{file=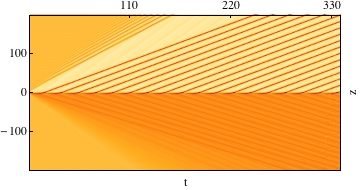,width=\columnwidth}
\caption{  Pseudo-color density plot $\rho_i(z,t)=\left|\psi_{i}  (z,t) \right|^2$ of component $i$ of an uniformly distributed Bose gas with constant interaction $g =1=g^R$ at $t=0$ evolving in time as a change of $g^L/g^R =1.9$ in self-interactions and $g^L_{ij} =0.9$ in cross-interactions has been implemented for $t>0$. Brighter areas correspond to higher condensate densities.\label{ho}}
\end{figure}
Fig. $\ref{ho}$ illustrates the evolution of dark soliton trains generated by the quantum piston scheme \cite{mynote}. The initial distribution has been specified by $n_0=1$ with $g_1=g_2 =1$ and $g_{ij}=0$ and we have changed self-interactions in both components by $g^L_i/g^R_i = 1.9$ as well as we have set  cross-interactions for both components  to be $g^L_{ij}=0.9$ on the l.h.s. ($z\leq0$) and $g^R_{ij}=0$ on the r.h.s..  In contrast to the single component case where dark solitons emerge once self-interaction ratios reach $2.2$ \cite{dark} dark soliton trains are generated due to the presence of cross-inetarctions. As both components $A$ and $B$ have the same density distribution,  $|\psi_1|^2 \simeq |\psi_2|^2$, Fig. \ref{ho} represents both of them.

Adding cross-interactions of the form $g^R_{ij} = c^R = 0$ and $g^L_{ij}  = c^L + c^{\text{up}}_{\text{crit}} g^R_i \geq 0$  the {\it effective} `self-interactions' for both atom clouds $s^{\text{eff}}$ are identical or increased at $z \leq 0$ and we have
\beq\label{hahaha}
s= \frac{g^L_i}{g^R_i} \leq  s^{\text{eff}} = \frac{ g^L_i+\sum^N_{j} g^L_{ij}}{g^R_i+  \sum^N_{j} g^R_{ij}}  = \frac{g^L_i + c^L+ c^{\text{up}}_{\text{crit}} g^R_i}{g^R_i} 
\eeq
for $N=2$ and $i\neq j$.
We have included $c^{\text{up}}_{\text{crit}} g^R_i$ within the numerator since solitons occur once a ratio of $c^{\text{up}}_{\text{crit}}$ in change of self-interaction strength is surpassed. Consequently for $g^L_i , c^L > 0$ an increase in frequency of the emerging soliton train can be listed by the increasing monotonicity of $r$ in Fig. \ref{freq}.  
Let us associate with the real number $g^L_i \geq 0$ the real number we want to manipulate and with $c^L \geq 0$ that one we want to add. For a set of arbitrary positive parameters $a \leftrightarrow g^L_i$ and $b \leftrightarrow c^L$ the logic is that adding $a$ to $b$ implies a number $c \geq a,b$ as the corresponding frequency of the soliton trains in components $A$ and $B$ is equal or higher than it would be only with the cross-interaction $c^L$ or self-interaction term $g^L_i$. Considering the logical structure of $\eqref{hahaha}$ the number $c$ can be associated to the outcome of an addition of two input numbers $a$ and $b$ if we just set $a = g^L_i$ and $b = c^L$. 

Furthermore, multiplication of two natural numbers $M,N \in \mathbb N^+$ can be implemented by considering a system of $N$-component condensates in similar states $|\psi_1|^2 \simeq |\psi_2|^2 \simeq \ldots$ and turning on cross-interactions between them. Setting for example ${g^L_i} \to g^L_i  + c^{\text{up}}_{\text{crit}} g^R_i \geq 0$ in \eqref{step} and $g^L_{ij} = g^L_i =M \geq 0$ while $g^R_{ij} =0$ in $\eqref{puss}$ yields
\beq\label{multikulti}
s^{\text{eff}} =  \frac{{g^L_i}' + \sum^N_{j \neq i} g^L_{ij}}{g^R_i} = \frac{M \cdot N+c^{\text{up}}_{\text{crit}}  g^R_i}{g^R_i} .
\eeq
Here again the frequency of emerging dark soliton trains is increased. Setting $N=a$, and $M=b$  the algebraic properties of closure, commutativity, the existence of an identity element, associativity of multiplication for natural numbers and no zero divisors are satisfied. (We note that $M$ can also represent a real number.) The corresponding properties apply to above addition scheme as well and implementing associativity for addition of three numbers is done by considering a three component condensate obeying a change in interactions of the form
\beq\label{three}
s^{\text{eff}}=  \frac{g^L_i + \sum^3_{j \neq i} g^L_{ij} + c^{\text{up}}_{\text{crit}} g^R_{i}}{g^R_i}.
\eeq
Writing $g^L_i  = g^L_{ii} $ and by associating $g^L_{11} = g^L_{23} = g^L_{32}  \leftrightarrow a $,  $g^L_{12} = g^L_{21} = g^L_{33}  \leftrightarrow b $ and the remainder analogously to $c$ implements a sum of three numbers that is associative.

To include the case of multiplication of a positive real number by a positive real scalar one can proceed as follows. Considering a two component system we set $g^L_{ij} = c g^L_i + c^{\text{up}}_{\text{crit}} g^R_i$, $g^R_{ij} =d g^R_i $ and by rescaling $c^{\text{up}}_{\text{crit}} \to \frac{(1+ d)}{(1+ c)} c^{\text{up}}_{\text{crit}}$ we obtain
\beq\label{mult}
 s^{\text{eff}} = \frac{(1+ c)}{(1+ d)}  \cdot \frac{g^L_i }{g^R_i } +  \frac{ c^{\text{up}}_{\text{crit}} g^R_i }{g^R_i }.
\eeq
Here the parameters $c$ and $d$ are not uniquely defined by the factor of multiplication, but together sufficient to represent in principle any positive real number. An analogue relation between change in self-interactions and frequencies of soliton trains holds for self-interactions being switched from entirely positive to negative values within a fraction of the condensate \cite{dark}. An illustration of this relation between change in interactions and frequency can be found in Fig. \ref{freq} and is represented by the graph $ra$. 

We note that the key action of inversion of a number, $1/x$, can be achieved in a two component condensate by associating it with change $s^{\text{eff}}$ inversely via $s \leftrightarrow 1/a$: Given a range of the function $1/x$ by $[k_1, k_2]$ we define
\beq
 s^{\text{eff}}  = \frac{d_1+ c^{\text{up}}_{\text{crit}} g^R_i }{g^R_i } \leftrightarrow \frac{1}{k_1}
\eeq
for a fixed value $g_i^L = d_1$. As we reduce $g_i^L \to d_2 < d_1$ we monotonically reach the smaller number $ \frac{1}{k_2}$, hence establishing a bijective association.

For $N$-component condensates a multiplication between a number $ -M$ and a positive natural number $N$ can be implemented analogously as in $\eqref{multikulti}$ by implementing self- and cross-interactions such that
\beq\label{negat}
s^{\text{eff}} =   \frac{ - M \cdot N +c^{\text{down}}_{\text{crit}}g^R_{i}}{g^R_{i}} .
\eeq

\section{    Computing method B} Next we consider a different computing method utilizing single component BEC (N=1). Formally
\eqref{multi} becomes
\begin{equation}\label{sing}
 i \frac{\partial}{\partial t}  \psi= - \frac{\partial^2}{\partial z^2} \psi  + g |\psi|^2 \psi.
\end{equation} 
Here $g=g(z)$ is of the form $\eqref{step}$. For atomic BEC a stepwise interaction function can be implemented by the means of Feshbach resonances \cite{fesch,dark}. As complementary mechanism we propose the implementation of  phase imprinting techniques  \cite{burg,burg2,dbec,pow}. As shown in \cite{dbec} phase engineering enables us to locally imprint a phase on the condensate wave function, i.e., $\psi \to \psi \cdot e^{i \phi (z)}$ with $\phi (z)$ being specified and highly controllable by the parameters of the experimental setup. By an appropriate choice of parameters this induces the generation of solitons, which experimentally are clearly observable \cite{pow}. In order to implement our computing scheme we imprint a phase over time, $\psi \to \psi \cdot e^{i \theta (z) t/2}$, of the simple form
\beq\label{stepph}
\theta(z) =  \begin{cases} \theta^R=c_5, & z>0 \\ \theta^L =c_6, & z \leq 0, \end{cases}
\eeq
where $c_5$ and $c_6$ denote fixed constants. 
Hence $\eqref{sing}$ becomes
\begin{equation}\label{sing2}
 i \frac{\partial}{\partial t}  \psi= - \frac{\partial^2}{\partial z^2} \psi  + g |\psi|^2 \psi + \theta \psi.
\end{equation} 
Here the phase $\theta = \theta (z)$ is utilized for generating soliton trains comparable to variable self- or cross-interactions. Numerical simulations of the condensate wave function show that a condensate initially specified by $\psi=\sqrt{n_0}=1$ and $g=1$ supports stable soliton trains above a threshold depending on the parameters of the system and in particular changing self-interactions $g \to g(z)$ at the same time implies an increase in frequency of the emerging soliton train, which in turn imposes an order structure on associated data, i.e., input number $a \geq 0$ associated with a change in self interactions $s = g^L/g^R \geq 1 = g^R = g$ combined with $b \geq 0$ associated with an appropriately chosen imprinted phase on the l.h.s. implies an output number $c \geq a,b$.

 \begin{figure}
\epsfig{file=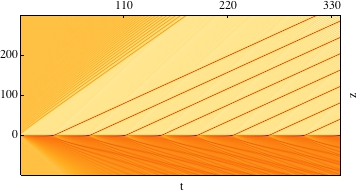,width=\columnwidth}
\caption{  Pseudo-color density plot $\rho(z,t)=\left|\psi  (z,t) \right|^2$ of an uniformly distributed Bose gas with constant interaction $g =1=g^R$ at $t=0$ evolving in time as a change of $g^L/g^R =2.2$ in self-interactions  has been implemented for $t>0$. Brighter areas correspond to higher condensate densities.\label{self}}
\end{figure}
 \begin{figure}
\epsfig{file=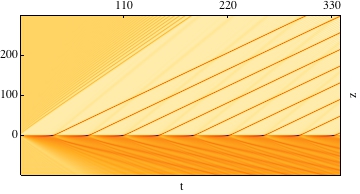,width=\columnwidth}
\caption{  Pseudo-color density plot $\rho(z,t)=\left|\psi  (z,t) \right|^2$ of an uniformly distributed Bose gas with constant interaction $g =1=g^R$ at $t=0$ and $\theta=0$ evolving in time as a step-like phase $\theta^L =0.7$ and $\theta^R=0$ has been implemented for $t>0$. Brighter areas correspond to higher condensate densities.\label{phase}}
\end{figure}
 \begin{figure}
\epsfig{file=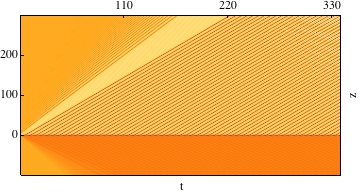,width=\columnwidth}
\caption{  Pseudo-color density plot $\rho(z,t)=\left|\psi  (z,t) \right|^2$ of an uniformly distributed Bose gas with constant interaction $g =1=g^R$ and $\theta=0$ at $t=0$ evolving in time as a change of $g^L/g^R =2.2$ in self-interactions and an additional step-like phase $\theta^L =0.7$ and $\theta^R=0$ has been imposed for $t>0$. Brighter areas correspond to higher condensate densities.\label{both}}
\end{figure}

 Fig. \ref{self} shows an example for a numerically generated soliton train evolving within a BEC generated solely via a change in self-interactions (see also \cite{dark}). Here we have imposed on an initial condition given by $\psi= \sqrt{n_0}=1$ with $g=1$ and $\theta = 0$ a change in self-interactions given by $g^L = 2.2$ and $g^R=1$. In Fig. \ref{phase} we show that introducing on the same initial condition a phase $\theta^L=0.7$ and $\theta^R=0$ generates as well a soliton train.   Imposing simultaneously a change in self-interactions and a step-wise phase specified by $g^L = 2.2$, $g^R=1$, $\theta^L=0.7$ and $\theta^R=0$ creates a distribution represented in Fig. \ref{both}. This density plot in particular shows that both soliton generating mechanisms in place yield a soliton train of higher frequency providing an order structure for the associated data. Finally we remark that in general implementing mathematical operations (such as $+$,$-$, etc.) depends on the form of $\theta(z)$ (and $g(z)$), which can be modeled to a very high degree using an appropriate setup of experimental parameters \cite{dbec}.

\section{   Conclusions} This letter has been devoted to present a new application for atomic BEC in the Gross-Pitaevskii regime and analogous systems as analog (classical) computing devices. We introduced two distinct and novel computing schemes utilizing Feshbach tuned scattering lengths and state-of-the-art phase imprinting techniques, which for massive BEC both are well-studied and experimentally tested. We showed how to utilize these concepts for computing numbers and gave various examples of implementations of different fundamental mathematical operations in single and multicomponent systems. 

\section{    Acknowledgments}
FP has been financially supported through his EPSRC doctoral prize fellowship at the University of Cambridge and by the King Abdullah University of Science and Technology (KAUST)
Award No. KUK-I1-007-43. I am very grateful for discussions with Hugo Flayac and Natasha Berloff.


\begin{thebibliography}{10}
\bibitem{mynote}
To follow the time evolution of $\psi$  induced by $\eqref{multi}$ computationally we have used a fourth order finite difference scheme in space together with a fourth order Runge-Kutta discretization in time.
 In simulations, we used $\Delta x = 1/6$, $\Delta t=0.0015$,  $z\in (-500,500)$ and absorbing boundary conditions.
\end{thebibliography}
  \end{document}